%% file: skycloud.tex
\documentclass[12pt]{article}
\usepackage[margin=1in]{geometry}                
\usepackage{times}                  
\usepackage{authblk}
\usepackage{listings}
\usepackage{xcolor}
\usepackage{graphicx}
\usepackage{url}

\DeclareFixedFont{\ttb}{T1}{txtt}{bx}{n}{8} 
\DeclareFixedFont{\ttm}{T1}{txtt}{m}{n}{8}  

\usepackage{color}
\definecolor{deepblue}{rgb}{0,0,0.5}
\definecolor{deepred}{rgb}{0.6,0,0}
\definecolor{deepgreen}{rgb}{0,0.5,0}

\title{A Cloud in the Sky: Geo-Aware On-board Data Services for LEO Satellites}
\author{\small Thomas Sandholm, Sayandev Mukherjee, Bernardo A Huberman}
\affil{Next-Gen Systems,
CableLabs, Santa Clara, CA}

\begin{document}
\maketitle

\input{abstract}
\input{introduction}
\input{overview}
\input{relatedwork}
\input{blockchain}
\input{leo}
\input{protocol}

\input{simulation}
\input{evaluation}
\input{conclusion}

\bibliographystyle{plain}
\bibliography{related}

\end{document}

%% file: abstract.tex
\begin{abstract}
We propose an architecture with accompanying protocol for on-board satellite data infrastructure
designed for Low Earth Orbit (LEO) constellations offering
communication services, such as direct-to-cell connectivity. Our design leverages the unused or under-used computing and communication
resources of LEO satellites that are orbiting over uninhabited parts of the earth, like the oceans. 
We show how blockchain-backed distributed transactions can be run efficiently on this
architecture to offer smart contract services.  

A key aspect of the proposed architecture that sets it apart from other blockchain
systems is that migration of the ledger is not done solely to recover from failures.
Rather, migration is also performed periodically and continuously as the satellites
circle around in their orbits and enter and leave the blockchain service area.  

We show in simulations how message and blockchain processing overhead can
be contained using different sizes of dynamic geo-aware service areas. 
\end{abstract}

%% file: introduction.tex
\section{Introduction}\label{sec:introduction}
With so many connected devices and their improved capabilities, the need for high-bandwidth connectivity
anywhere and anytime keeps growing at an exponential rate.
As a result, high-population areas usually see investments in high-capacity communication infrastructure
such as cell towers, small-cell antennas, radio heads and base stations.
For an end-user device wanting access to a wireless mobile network, having low-latency access to this infrastructure not only allows the initial
connection to be established faster, but also encounters fewer restrictions on the effective throughput, given timeouts of various acknowledgments in the protocols.

For remote or sparsely populated areas not served by the communication infrastructure described above, satellite communication is an attractive, albeit expensive, solution. Recent improvements in antenna technology as well as satellite launch economics
have contributed to a 12-fold increase in the number of Low-Earth-Orbit (LEO) satellite launches
in the last decade~\cite{vanelli2024}. Because LEO satellites orbit the earth at a lower
(and fixed) altitude they are ideal for providing communication services.

Current Non-Terrestrial-Networks (NTN) follow a \emph{bent-pipe} architecture whereby the
satellites serve as a simple relay between communicating parties on earth (the user device and a terrestrial base station, say). As a result,
two-way communication between a target device and a base station providing
network services result in messages traveling between the LEO satellite and earth four
times, severely impacting both the latency and bandwidth the networks can provide.
To overcome this overhead, recent efforts to move some core network services on-board
the satellites have gained in popularity. The resulting architecture is often
referred to as a non-transparent or regenerative, indicating that the satellite 
takes an active part in the communication beyond just relaying between parties on earth.

Apart from latency and bandwidth concerns, NTN networks also have to address the challenge of
spectrum access. As LEO satellites circle the earth they need to make sure their transmissions
do not interfere with (typically nationally licensed) terrestrial networks anywhere in their orbit,
while at the same time being easily accessible to standard communication devices, such as cell phones.
This has lead to a natural collaboration between Mobile Network Operators (MNOs) and
Satellite Network Operators (SNOs). While a LEO satellite can circle the earth in 90 minutes, its orbital arc typically allows it to serve a given earth-located station for only about 5 minutes. Thus, any connectivity between a user device and a terrestrial base station through an LEO satellite needs to be established efficiently,
depending on the current local constraints and spectrum availability.

In this paper, we propose to host a bandwidth ledger that enables on-demand
purchases of cellular bandwidth akin to the way compute resources are
purchased in the Cloud, while allowing revenue sharing between the MNOs and SNOs.
Given the latency concerns described above, we further propose to host this ledger entirely
on-board the LEO satellites, using their Inter-Satellite-Link (ISL) free-space
optical communication links. 

In Section~\ref{sec:overview}, we summarize the principal contributions of the present work.  We then provide an overview in Section~\ref{sec:relatedwork} of related work, discussing the differences between our approach and those of other researchers.  In Sections~\ref{sec:blockchain} and~\ref{sec:leo} we provide a quick background overview of the two main technologies that are part of our proposed solution, namely Blockchains and LEO satellite constellations.  The main section describing our proposed protocol is Section~\ref{sec:protocol}, followed by Section~\ref{sec:simulation} describing a simulator and visualization tool for on-board blockchain processing in LEO constellations.  We present the results of evaluation of the proposed protocol in Section~\ref{sec:evaluation} and our conclusions in Section~\ref{sec:conclusion}.

%% file: overview.tex
\section{Contributions}
\label{sec:overview}
Note that communication between satellites is literally happening at the speed of light, but computational overhead can be an
issue, in particularly with re-generative workloads or multi-tenant radio heads serving many MNO spectrum
bands (e.g. on-board gNodeBs). Moreover, adding capacity to these networks is costly, as the circular motion and
equal spacing means that there are the same number of satellites serving low-traffic areas such as big
oceans and deserts, as the most populous areas such as metropolitan cities.

Our solution addresses all these challenges by running a distributed ledger to execute smart contracts
using the unique structure of LEO ISL communication as well as the geographic properties of orbital
cycles.

We propose two data infrastructure primitives: a {\it gossip protocol}, and a {\it distributed transaction
processing protocol}, that are customized to be efficient in LEO communication ISL networks and can be used
to implement smart contracts as well as distributed databases for core communication services to avoid a round-trip
to ground stations. We introduce the notion of a Service Area (SA) which denotes a geographical area
where data services will be actively hosted. 
We further introduce the novel concepts of {\it leader row} and {\it neighbor migration} to support moving
the cluster of active participating satellites in and out of the service area as they move around in their orbits.  

%% file: relatedwork.tex
\section{Related Work}\label{sec:relatedwork}
The many challenges of offering cellular connectivity from satellites, including doppler effects and latency issues,
are described in~\cite{guidotti2017}. These and other challenges in using the traditional satellite
communication spectrum in the L-Band and S-Band for direct-to-cell
connectivity~\cite{pastukh2023} are the reason for SNOs and MNOs to coordinate their usage and allocation of spectrum resources. 

The FCC has also recognized the value of
re-using MNO spectrum for extended or supplemental coverage through satellites\footnote{\tiny \url{https://www.fcc.gov/document/fcc-proposes-framework-facilitate-supplemental-coverage-space-0}}.

Previously, we have designed smart contracts for more efficient spectrum sharing
among terrestrial MNOs~\cite{sandholm2023,sandholm2021,sandholm2021b,sandholm2020}. The 
present work extends this idea to LEO-based networks.

A similar approach of micro-contracts between MNOs and LEO SNOs is described
in~\cite{liu2024}, where the authors propose an architecture for giving MNO customers easy access to
SNO-provided multi-tenant direct-to-cell networks via standard SIM cards. Customers purchase
tokens to use satellite services from their MNO, which will be written into trusted hardware on their
SIM cards. When satellite access is needed, the SNO will be able to verify the validity of the token
as well as mark it as used before providing service.

We have previously envisioned a similar architecture for ad-hoc multi-provider bandwidth purchases using eSIMs and an open
distributed ledger market in~\cite{sandholm2023} and~\cite{sandholm2021} to overcome the
inefficiencies of establishing new roaming contracts between providers. The advantage of our approach compared to
that of~\cite{liu2024} is that contracts do not need to be negotiated ahead of time and there is no need
to keep tokens on SIM cards. Instead, our approach requires the presence of a distributed ledger accessible
to both the end-user device and the mobile operator, and thus the performance of the system depends on the placement and migration of the distributed ledger. 

QoS optimization is addressed via edge-service placement in LEO constellations in~\cite{pfandzelter2022}, where the authors select an optimal subset of satellites to host a specific service like a CDN
or IoT data processor. In the scenario they address, the primary benefit is cost as
only a subset of satellites deemed optimal to host services will be equipped with the necessary
hardware. In contrast, in the present work we propose a software allocation solution that reduces the
load on satellites to allow lower capacity hardware to be provisioned on all satellites. 
Moreover, our software placement is aware of the geographic position of the satellites in order to exploit
idle resources on satellites not actively engaged in communication services.
Our focus is less on reducing hops and instead on limiting broadcast overhead while still getting the benefit of replication.
We note, however, that we chose to adopt their +GRID 2d torus model of ISL communication for our work as well. 

Geo-aware LEO ISL routing schemes are investigated in~\cite{roth2021}. The key to their approach
is to embed geographical information in the MAC addresses of the communicating terminals to route
packets more effectively without having to change the allocated IP addresses. With this approach
they can better handle handovers and reduce delays in the dynamic and constantly moving network
conditions of LEO satellites. 
Our focus is more on data infrastructure service provisioning and
load balancing while taking geo-position into account, as opposed to routing. 

ISL routing is also considered in~\cite{sun2002}, where the main goal is to effectively find
alternative paths in case of failure while making sure that the link capacity in the network as a whole is
optimized. They argue that path restoration instead of link restoration is a more efficient way of
recovering from link failures. We only indirectly deal with failures in that we process transaction
on a large set of satellites concurrently and in virtual synchrony, so that reaching any of these
satellites would allow access to the up-to-date data with eventual consistency guarantees. 

An auction-based mechanism to improve channel efficiency between ground stations and
LEO satellites was proposed in~\cite{cheng2023}. Our bandwidth market can also offer
auctions as demonstrated in~\cite{sandholm2021}, but is more focussed on contracts 
between end-users and MNOs via SNOs. Furthermore, the work presented here is about making best use
of inter-satellite links for on-board state management as opposed to ground station communication.

There have been many efforts to simulate satellite communication in software, e.g. \cite{schubert2022}, \cite{pfandzelter2022}, \cite{kempton2020}
and \cite{puttonen2015}. We opted to build our own simulator to focus on the parts that mattered to our study,
namely the LEO ISL links and the service migration and movement of satellites, as well as the ability to run custom software on
each satellite node. The way we implement orbital planes as processes and satellite nodes as threads 
allows us to simulate large constellations effectively while easily facilitating and monitoring all communication between nodes.
Furthermore, our simulator is written entirely in Python3 with only a couple of external dependencies
for inter-process communication (HTTP REST). The visualization is written in standard JavaScript HTML5 running
on a local Web server, which makes it easy and quick to set up and run locally on any laptop, as well as to demo remotely.

%% file: blockchain.tex
\section{Blockchains and Distributed Ledgers}\label{sec:blockchain}
Blockchain technology was popularized with the virtual bitcoin currency~\cite{nakamoto2008}
and provides a way to maintain a distributed ledger consistently and scalably
across a large number of nodes. The 2-phase-commit transaction
protocols used in traditional relational databases~\cite{bernstein2009}, for instance, scale very poorly
as the transaction as a whole fails if only one party maintaining state fails. Blockchains originally
provided a way to ensure consensus by what is referred to Proof of Work (PoW) where
only parties that solved complex math problems were allowed to write into the ledger.
The way a blockchain serializes blocks and maintains a hash of the previous block
in the hash of the current block, it is easy to validate the internal consistency of the chain.
PoW blockchains may contain competing forks of the blockchain but the longest chain wins
if there is a conflict. These types of blockchains are appropriate in environments where
anyone is allowed to write into the ledger when partaking in a large community of untrusted parties.
The computation to solve math problems is however energy hungry and since LEO satellites
have a limited lifetime depending on how long their battery lasts, this type of blockchain,
referred to as permissionless, is not appropriate for our use case.

Instead we focus on permissioned blockchains that only allow authenticated and trusted
participants to write into the ledger. Hence the blockchain can never fork into
competing branches and no work is wasted. Some nodes may have fewer blocks but will
eventually catch up.  Still, the protocol needs to deal with failures while ensuring
a consistent ordering of transactions. This is typically done by having a single node act
as the leader to provide consistent ordering. Electing a leader is a critical part of the protocol,
as having no leader or multiple ones will cause the system to fail as a whole. Leader election
can be done by various consensus algorithms such as Paxos~\cite{lamport2001} or Raft~\cite{ongaro2014}. As we shall see later we
can provide a more suited and simpler leader election algorithm for the LEO case that
is similar to the token ring leader election algorithm~\cite{shirali2008}.

Permissioned blockchains like Hyperledger Fabric\footnote{\url{https://www.lfdecentralizedtrust.org/projects/fabric}}
go through the following high level steps
to process a transaction: (1) execute the transaction against local state to ensure validity
and record all versions read and what is written in each atomic unit (transaction), (2) send the
read and write operations to the current leader to order all transactions globally, (3) once a sequence
number is attached to each transaction or optionally they have been bundle in larger blocks
broadcast them back (via a gossip protocol)  to all nodes maintaining the blockchain state, 
(4) nodes receiving the transactions will validate and execute them in the order given to a local blockchain
and state repository. Validation ensures that the version read has not been updated by some other transaction
before writing to the same state.

As we shall see later we follow the same high-level steps which could be compared to the virtual synchrony~\cite{birman1987}
state replication approach, but with customizations at each step.

This process allows reads and writes to be done with eventual consistency~\cite{vogels2009}, 
or optimistic locking as opposed to with strict atomicity, consistency, isolation
and durability (ACID) guarantees, known from protocols such as two-phase commit~\cite{bernstein2009}. 
As previously mentioned, the issue with the ACID protocols is that they scale poorly across large sets of replicas 
as failures become more likely.

A blockchain could in theory be implemented on a single node to ensure consistency but that would also
reduce availability, so most deployments have the architecture of a distributed ledger with eventual
consistency guarantees. This is also the model we follow.

Eventual consistency (guarantees) can be defined as follows. 
Assuming there are $n$ nodes in the service area, if $w$ nodes acknowledge they committed the transaction, and $r$
nodes are used to read the data, then consistency can be guaranteed iff $w+r > n$ (see~\cite{vogels2009}). The tradoff
follows from relaxing the C in the CAP theorem~\cite{brewer2000}.

%% file: leo.tex
\section{LEO Constellations}\label{sec:leo}
Low-Earth-Orbit (LEO) satellites are typically orbiting at a fixed
altitude above the earth in the range of 311 to 621 miles~\cite{cottin2017}. 
In contrast to GEO
satellites that follow the rotation of the earth to appear at a fixed point from
a given position on earth, the LEO satellites rotate faster than the earth spins.
The velocity and thus orbital period depends on the altitude
of the satellite and mandates the number of satellites needed for full coverage. 
As an example, a LEO satellite at 391 miles altitude orbits
the earth in about 97 minutes. A typical constellation at this altitude designed to
provide full coverage with 34 satellites in each orbital path~\cite{pfandzelter2022} 
results in a new satellite
appearing over a fixed point on earth every 3 minutes.

Satellites on the same orbital path are referred to as an orbital plane. They are placed at equal distance from each
other with a wraparound, so the first satellites is at the same distance from the second as from the last.
These inter-satellite distances do not change over time. Orbital planes are further
organized into shells or constellations\footnote{Sometimes constellations refer to a set of shells
but we only assume intra-shell communication is available here so use constellation and shell
interchangeably.} where each plane has the same altitude over earth
and inclination angle\footnote{angle at which its path crosses the equator}. Communication across orbital planes are possible but the distance to the
nearest neighbor in a different plane may differ over time, although it is
predictable and cyclical. Hence, within plane communication is more reliable. 

\begin{figure}[htbp]
	\centering
        \includegraphics[scale=0.4]{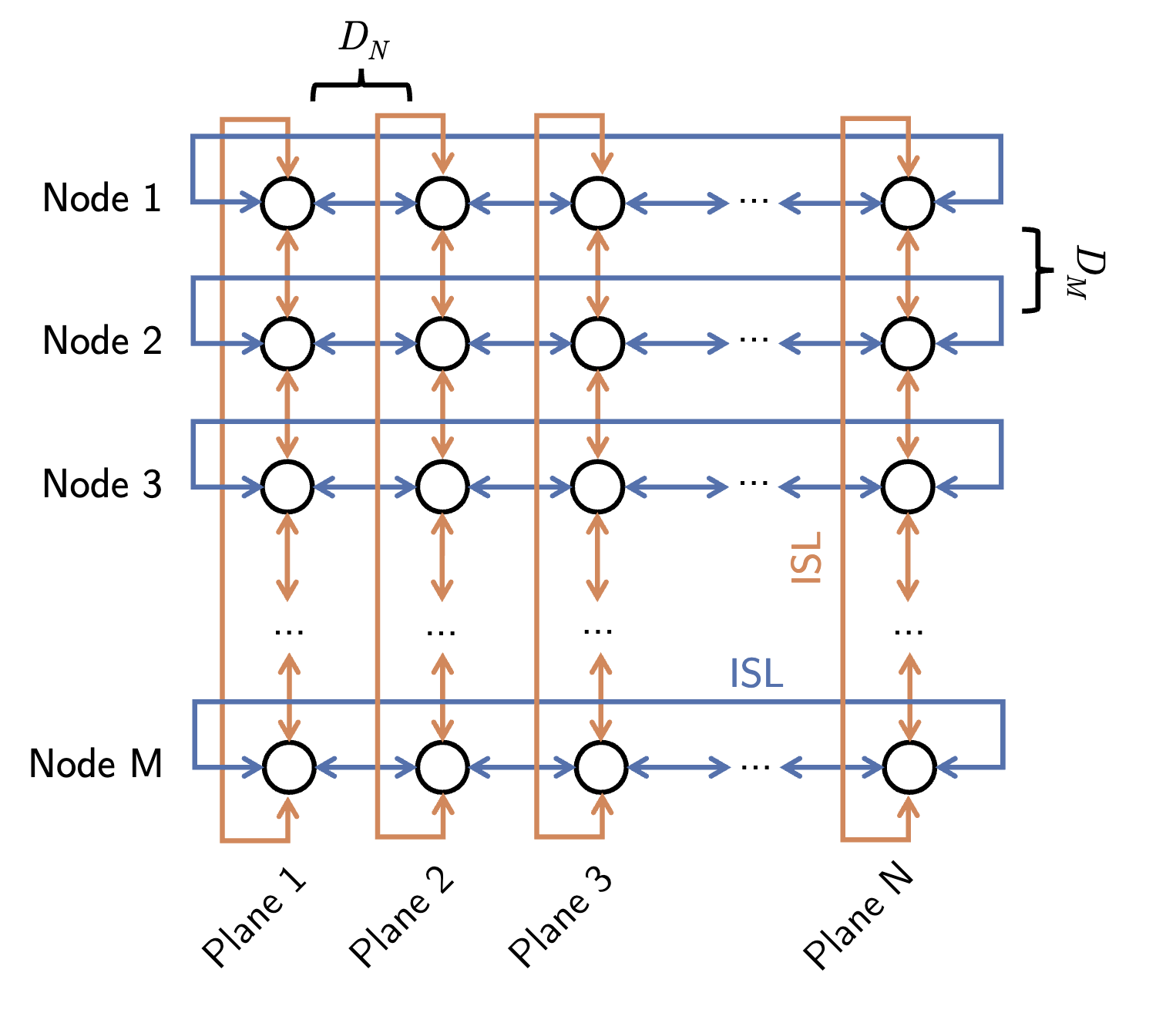}
        \caption{+GRID 2d torus ISL. Source: Pfandzelter and Bermbach~\cite{pfandzelter2022}.}
\label{torus}
\end{figure}

Each satellite  typically has 4 inter-satellite links to neighboring satellites, e.g. west, east, north and south, using free-space
optical connections to transmit at the speed of light. The network in a shell can thus be
seen as a connected mesh forming a 2d-torus where there is a wrap-around both in rows 
and columns (see Figure~\ref{torus}). This architecture is referred to as a +GRID inter-satellite link (ISL) 
network~\cite{pfandzelter2022}, 
and it is the one we assume for our work. 

As described in Section~\ref{sec:overview}, a service area (SA) is defined as an area where satellites are typically
less loaded with communication workloads, such as over the oceans. Any satellite passing over this area
will take an active part in hosting the data services. 

For example, the Pacific Ocean stretches about $15,500$km 
from the Arctic to the Southern Ocean and $19,800$km from Indonesia to the coast of Colombia, which is close to
half the circumference of the earth. So up to a third of the planes and half of all the satellites in a plane 
could hover over the Pacific Ocean at any given time.
A typical LEO constellation would make up about 22-28 orbital planes with 5-72 satellites per plane or 375-1584
satellites in total\footnote{These numbers are based on the data in Table I in~\cite{pfandzelter2022}}. Thus about 8 planes and 39 satellites per plane or about 321 satellites could be wasting
their capacity while idling over the Pacific. Having 300+ nodes in a distributed ledger with virtually
unlimited speeds to interconnect them is clearly a resource worth exploiting.

One major issue with these networks is that adding more capacity by sending up more satellites
to increase coverage also means that there is more waste in terms of periods where the satellites
are idle. Furthermore, these periods are predictable and cyclical thus easy to identify and exploit without any
central coordination\footnote{each satellite node can determine independently when they are above a certain
geographic area without a need for external communication}, which is the key idea behind our approach, which we describe next.

%% file: protocol.tex
\section{Protocol}\label{sec:protocol}
Next, we discuss the different parts of our protocol: gossip, transaction processing,
migration and smart contracts.

All our protocols rely on the 2d-torus architecture of LEO ISL. Furthermore we define
a blockchain Service Area (SA) as a block of contiguous nodes (each node being hosted on a satellite) over some geographic area with low terrestrial traffic load,
such as the Pacific Ocean.  All satellites in the constellation may
communicate with each other through the one-hop architecture of the 4 ISL links (see Figure~\ref{torus}),
but broadcasting is only done in the service area, and only nodes in the service
area take active part in processing transactions and maintaining up-to-date blockchain state.
Owing to their orbital movement, the set of nodes that are in the service area keeps changing, so we need
to perform continuous migrations of the blockchain ledger to nodes that move into the service area. Note that the nodes currently in the service area do not need to be informed about the new nodes moving into the service area, because the movement of every satellite is completely predictable. Thus, each node in the service area can independently determine when it will move out of the service area and which node will enter to take its place.

\input{gossip}

\input{transaction}

\input{migration}
\input{contract}

%% file: gossip.tex
\subsection{Gossip and Routing}\label{sec:gossip}
Updates on one node need to be broadcast efficiently to other nodes in the blockchain cluster to ensure
the window of inconsistency is kept small. At the same time a single path to distribute updates
can be a single point of failure, e.g. if all updates go through a central node. So, for reliability
most blockchain systems implement some form of the gossip protocol~\cite{demers1987}. These protocols are modeled after
epidemic virus infection spread in a human population: if infected, you then infect your nearest (uninfected) neighbors who go on to infect their (uninfected) neighbors, while an encounter between two infected people does
not propagate the virus further. Typically a handful of neighbor nodes are configured to propagate
gossip messages to each other, while avoiding to propagate to a neighbor that sent you the broadcast message.
A message that has already been seen, e.g. previously received by a different neighbor, is not propagated again. 

The gossip protocol is also efficient, in the sense that $N$ nodes can be synchronized in $O (\log N)$ iterations despite node failures and data losses in transmission~\cite{demers1987}. The gossip protocol on a torus (which is the scenario that matches LEO satellite constellations) was analyzed in~\cite{meyer2002}.  Improved gossip protocols designed specifically for use in blockchains were proposed and evaluated in~\cite{berendea2020},~\cite{korkmaz2022}, and~\cite{saldamli2022}.

In a LEO constellation architecture that follows  the +GRID ISL structure the neighbors are simply the 
north, south, east, and west connections. If a message comes from one direction it is propagated into the
other three, and previously seen messages are dropped. Furthermore we restrict propagation to
within the service area we defined, typically a grid but it could have any shape as long as all the nodes
in the service area are contiguous. All messages traversed also receive the gossip due to the one-hop
communication structure, ensuring the entire service area receives a broadcast efficiently.

Communication within an orbital plane tends to be more reliable and thus more nodes are reached within
a time unit if west/east spreading is prioritized, but propagation can be done concurrently in all directions. 

In some cases a node outside the service area may want to trigger a broadcast within the service area
or simply execute an operation on any node that is closest within the one-hop structure and at the 
same time is currently in the active service area.
Depending on where the broadcasting node is located and how the service area is configured the
shortest path to a target could be in the opposite direction using the wrap-around nodes of the 2d-torus.

For instance, a transaction may be created from any node but has to be executed in a node in the current service area. In that case the leader node that orders transactions then broadcasts the ordered transactions within the service area
only.

Changing the leader node could lead to disruption in transaction processing and hence we want to do
that as infrequently as possible. Thus a leader is not dropped until it leaves the service area. 
The leader's identity is also broadcast through the gossip protocol within the service area.
To optimize communication we limit the nodes that can be leaders to a single row, the {\it leader row},
i.e., the nodes in a selected orbital plane that are within the service area. In theory, it is sufficient for only the leader row nodes to know the identity of the leader, as all other nodes could send messages to the leader via the nodes on the leader row.
The fewer the nodes that need to know who the present leader
is, the smaller the window of downtime when a new leader is elected (details of the method of electing the new leader node are given in Section~\ref{sec:migration}). After the new leader node has been elected but before the current leader node has exited the service area, if the current leader node receives any messages intended for the leader, it can simply forward them to the new leader node.

%% file: transaction.tex
\subsection{Transaction Processing}\label{sec:transaction}
Transaction processing is at the core of our protocol as it is what
ensures eventually consistent distributed states and availability
even in the case of partial failures. A survey of blockchain consensus algorithms may be found in~\cite{hussein2023}. Distributing state across a large
area becomes even more important in an ISL network where only single node
hops are possible. As we previously mentioned it is enough to contact
any node (satellite) currently in the service area to read the current state.

State here is simply a key-value database where the key is a string and the value is 
an arbitrary object defined by the application, e.g., a JSON dictionary. The state is
updated with transactions, a transaction being defined as an ordered list of atomic read and write operations on the state. Each such operation either succeeds or fails. Transactions
can be submitted by any node in a constellation and executed by any node in the service
area. However, the order in which transactions are executed is kept consistent across
the constellation. 

The ground truth of the order of all transactions is logged in a blockchain ledger, replicas of which are
maintained by all the nodes in the service area. The content of a block in the blockchain is
the read and write operations for the transaction as well as the hash of the previous
block in the blockchain. Hence the validity of transactions can be independently verified
by any node in the service area. The state is maintained separately but logs the block id of
the last transaction that made an update to the state.  Whenever a key in the state is written
to, the version is bumped up and all read operations in transactions note which version they read.  Note that a transaction
can span many keys in the state, e.g., read from one state key and write into another.

The processing of a transaction is described in full below.

\begin{enumerate}
\item{A transaction is submitted by any node in the constellation. It is then routed via ISL to the nearest
service area prioritizing within-plane hops. Note that if a service area is defined to span the full set of orbital planes
this routing step only requires within-plane routing.}
\item{The node within the service area that receives the transaction executes it locally without writing to the ledger (blockchain) or state. This is done by reading and writing to a separate in-memory copy of the state to ensure the transaction
is valid assuming: (a) the node's state is up to date, and (b) the transaction does not rely on state that has since been updated,
i.e., there is no version mismatch between the read key and the key in the state of the node executing the transaction.}
\item{If local execution succeeds, the node forwards the transaction to the leader node for ordering.}
\item{The leader node attaches a sequence number (which is global to the constellation) and then broadcasts the transaction, or a set of ordered transactions, to all the nodes in the service area using the gossip protocol\footnote{This is similar to the Ordering service process described at \url{https://hyperledger-fabric.readthedocs.io/en/release-2.2/txflow.html}}.}
\item{A node receiving the transaction broadcast will validate the transactions in sequence and write the transactions
that succeeded to the blockchain as well as update the state and key version numbers accordingly. The same verification
as in the local execution of the transaction is performed with the difference that the validation is done in an order
mandated by the leader node as opposed to the time of arrival.}
\item{When submitting a transaction, a transaction ID is generated and the submitter may query any node in the service
area to check whether a transaction completed successfully. A successfully completed transaction including all its read
and write operations is then written into the blockchain and the state is updated accordingly.} 
\end{enumerate}

\begin{figure}[htbp]
        \centerline{\includegraphics[scale=0.4]{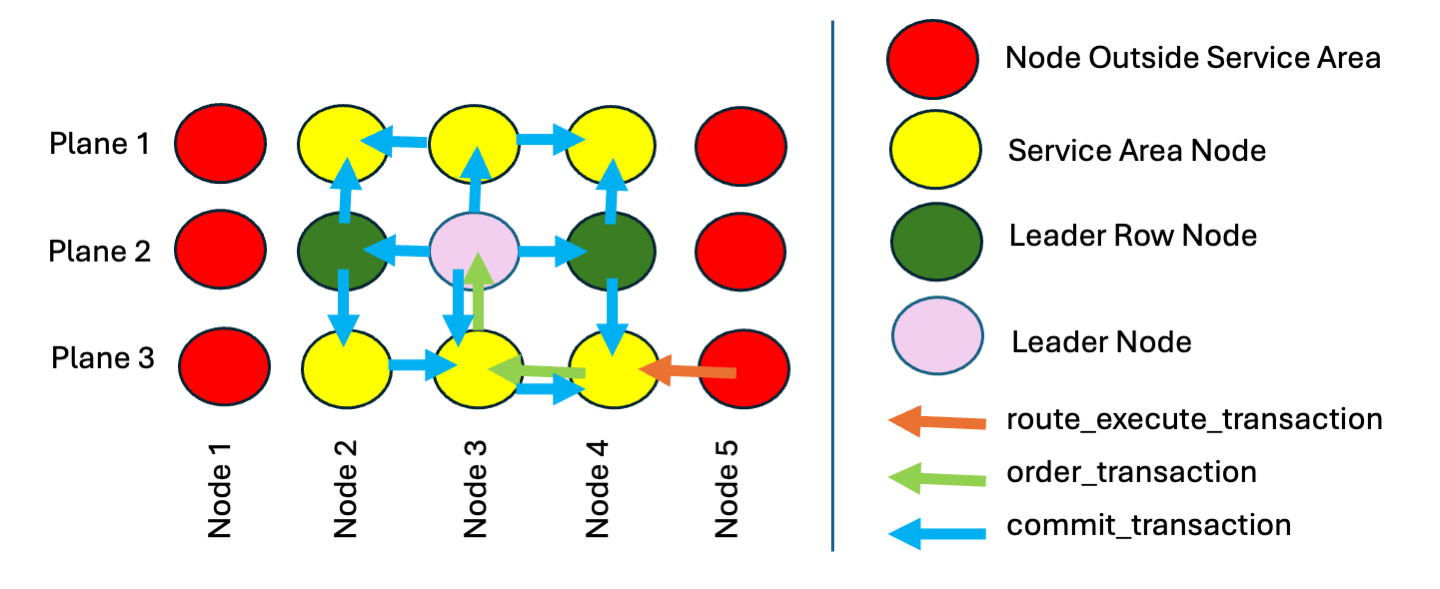}}
	\caption{Transaction route execute, order, and commit (gossip) message passing.}
\label{transaction}
\end{figure}

The messages passed during processing of a transaction can be seen in Figure~\ref{transaction}.

%% file: migration.tex
\subsection{Service Area Migration}\label{sec:migration}
They key aspect of our protocol that sets it apart from other blockchain
systems is that migration is not only done to recover from failures
but also done periodically and continuously as the satellites
circle around in their orbits and enter and leave the service area.
Therefore it needs to be very efficient and cause minimal disruption
to transaction processing.

The migrations are complicated by the fact that not only do we need
to make sure the state and the blockchain migrates properly but also that
the leader role migrates as well, which includes migrating 
in-memory state of the old leader to the new leader, and the election of a leader
using a consensus protocol.

As a set of nodes are about the leave a service area and another set is about to
enter the service area the following steps, which we refer to as {\it neighbor migration}, are performed:

\begin{enumerate}
\item{If the current leader is among the nodes that are leaving the service area, a new leader is elected. The old leader simply sends a message to the new leader that it should take over the role as leader. The leader election can only happen within a preconfigured orbital plane and thus the same row in the ISL torus. The new leader is the node furthest to the east in the leader row inside the service area. If that node does not respond the one west of it is elected and so on. A more formal ring leader election algorithm may also be executed. Once a new leader has been elected, the newly-elected leader broadcasts that it has taken up the role of leader within the service area using the gossip protocol.}
\item{Each node about to enter the service area synchronizes and updates their state and blockchain with their neighbor directly to the west already inside the service area. Only the blocks 
in the blockchain that were appended after the last rotation (of these nodes) in the service area need to be retrieved, and this communication is efficient as it is a single hop. Furthermore, all the nodes
across all orbital planes that are about to enter the service area can do this migration concurrently and using different target nodes to
synchronize from. This again makes the migration operation scalable and efficient.}
\item{Finally, when the migration is complete the west-most nodes in the service area orbits can drop out, and the borders of the
service area can be moved to include the nodes that just migrated into it. Note that the timing of these steps can be predicted within each node, based on the location of the nodes, and thus there is no need to send messages to trigger these steps.}
\end{enumerate}

In order for the new nodes (those just entering the service area) to execute transactions locally they need an up-to-date version of the state as well as the blockchain. For them to write into the ledger they also need to know the last known transaction sequence number so they can buffer
incoming ordered transactions if they arrive out of order. Finally, if a new node is the new leader, it also needs to have an up-to-date
version of the global sequence counter assigned to transactions that are to be ordered.

%% file: contract.tex
\subsection{Smart Contracts}\label{sec:contract}
In order to be able to test our transaction processing infrastructure
better, we also provide a smart contract programming
construct that allows us to define smart contracts.
A smart contract is simply an interface with methods
that read and write from the state maintained in the
blockchain, as well as define what the structure of
the state is. Each method when executed generates
a transaction comprising an ordered list of read and
write operations that can be submitted into the 
constellation for processing and that will be
written to the ledger with the eventually consistent
guarantees.

Below is an example of a smart contract defining a bank
account contract with the ability to transfer money
between accounts.

\noindent
\begin{minipage}{\linewidth}
\lstset{language=Python,basicstyle=\scriptsize\ttfamily,showspaces=false,showstringspaces=false,morekeywords={self},  
keywordstyle=\ttb\color{deepblue},
emph={MyClass,__init__},
emphstyle=\ttb\color{deepred},
stringstyle=\color{deepgreen},
frame=tb} 
\begin{lstlisting}[backgroundcolor = \color{lightgray}, framexleftmargin = 1em, framexrightmargin = 1em]
class Contract:
    def call(self, contract, op, args):
      self.transaction = Transaction(contract=contract)
      getattr(self, op)(**args)
      return self.transaction

class AccountContract(Contract):
    def create(self,balance=0):
      self.transaction.write(str(uuid.uuid4()), {"balance": balance})
    def transfer(self,from_account=None, to_account=None, balance=0):
      value, _ = self.transaction.read(from_account)
      value["balance"] = value["balance"] - balance
      self.transaction.write(from_account, value)
      value, _ = self.transaction.read(to_account)
      value["balance"] = value["balance"] + balance
      self.transaction.write(to_account, value)

def register(name, clazz):
  contracts[name] = clazz

register("AccountContract",AccountContract)

## Example Usage:  
## transaction = invoke_contract("AccountContract","transfer",
##                               {'from_account': account1, 'to_account': account2, 'balance':2})
def invoke_contract(contract, op, args):
    return contracts[contract]().call(contract,op,args)
\end{lstlisting}
\end{minipage}

%% file: simulation.tex
\section{Simulation and Visualization}\label{sec:simulation}
To be able to test transaction processing and smart contract execution
in a constellation in motion, we developed a simulation using Python and
a Web visualization.

In the simulator we define ISL communication paths such that each node can only communicate with its immediate neighbors to the north, south, west and
east. Each node has a $(x,y$) grid coordinate
representing satellite $x$ in an orbital plane $y$. A service area defines a range
of $x$ values and a range of $y$ values that may be updated at any point in the
simulation to account for orbital movements.

Since a constellation may have thousands of nodes that can all communicate with each other
concurrently and independently, we designed the simulator to use as few resources as possible
while still being able to run realistic transaction processing scenarios and scale up
to large constellations. Each orbital plane is implemented as a separate Python process
that exposes a REST API to communicate with other orbital planes (in north-south links).
Inside an orbital plane (west-east links) all communications are done with a thread pool.
In general a good size of the thread pool is the number of nodes in the orbital plane,
but it can also be set based on load.

Each node writes to its own copies of the state and blockchain, both of which are represented by JSON files in the simulator.
When a node wants to send a message to another node it simply specifies the $(x,y)$ coordinates of the
target and the simulator will use a combination of one-hop messages within the thread pool and REST
APIs to reach the target using the routing and gossip protocol defined previously.

At any time, a node can be asked to synchronize its state with its neighbor to the west.
Similarly any node can be asked to assume the role of a leader. Nodes within the current
service area can execute and validate transactions locally and submit validated transactions
to the leader node for ordering. The leader node can sequence transactions and broadcast
the sequenced transactions to all service area nodes using the gossip protocol described previously.

For evaluation purposes we define a smart contract that can create accounts and transfer money between accounts, as well as monitor
the balance of each account as the satellites move around their orbits.

A web simulator shows a grid of the 2d-torus with the current state in each node for a given
account. It also shows the current nodes in the service area (yellow), the nodes outside the
service area (red), the nodes in the leader row (green) and the current leader (pink).

The web simulation moves all satellites one step west every time interval (configured to 10s).
At any time, transactions may be executed, such as creating new accounts and transferring money
between accounts. The service area follows the rotation of the earth to always reside above
the Pacific Ocean.

We use a demo constellation of 4 orbital planes and 28 satellites per plane, where 6 satellites
across all 4 planes cover the service area at any given time.

\begin{figure}[htbp]
	\centering
        \includegraphics[scale=0.11]{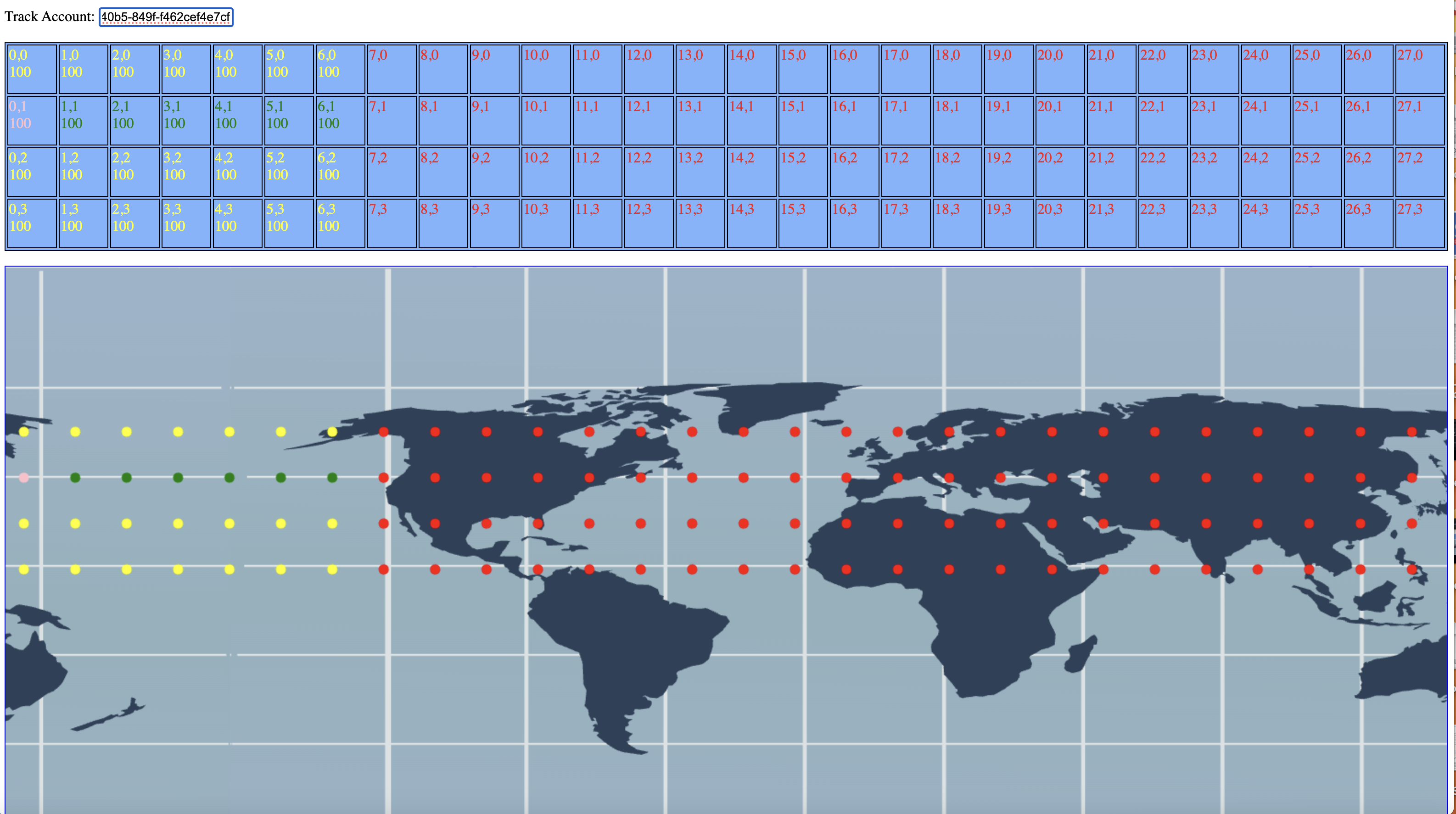}
        \includegraphics[scale=0.11]{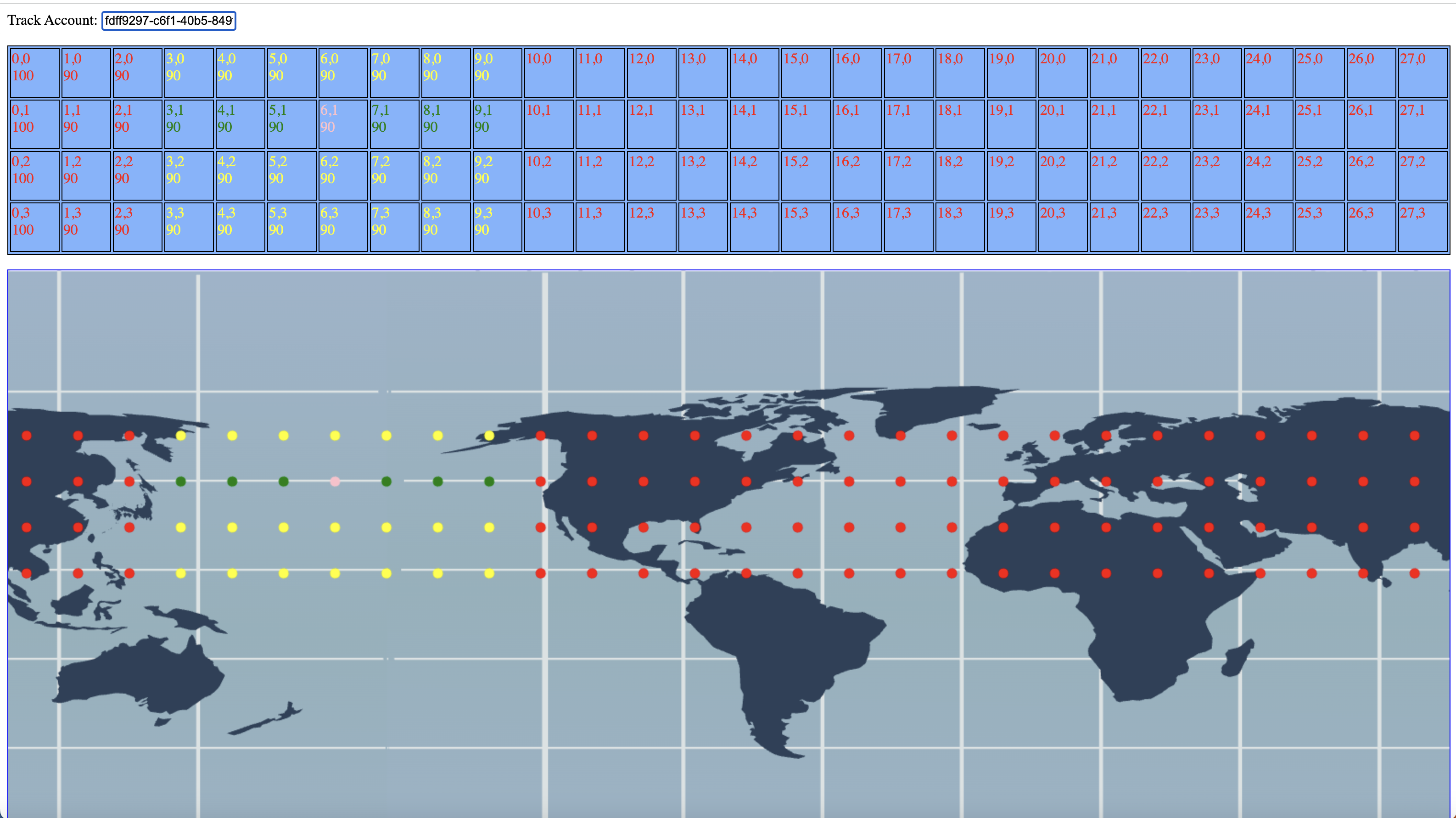}
        \includegraphics[scale=0.11]{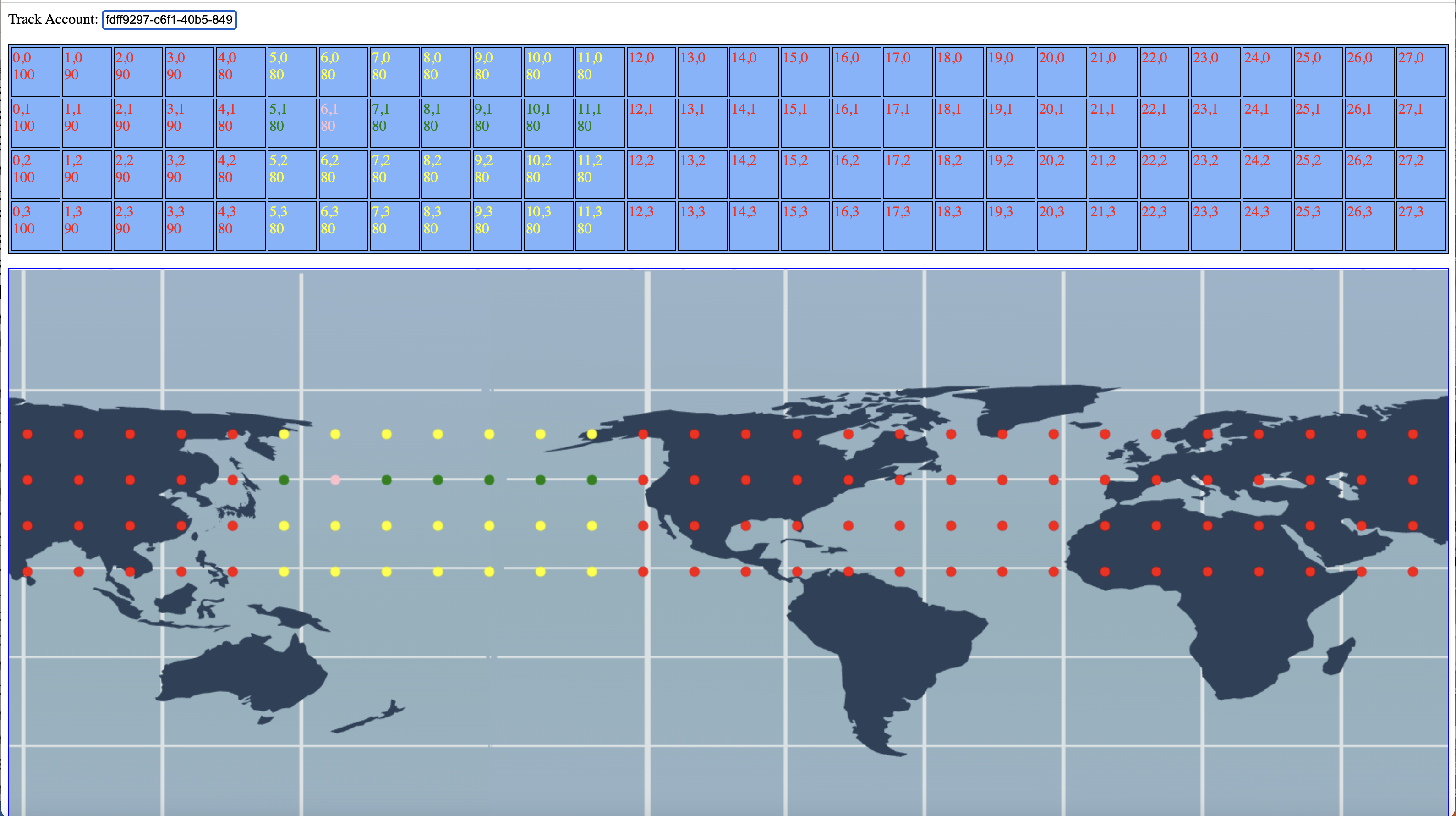}
        \includegraphics[scale=0.11]{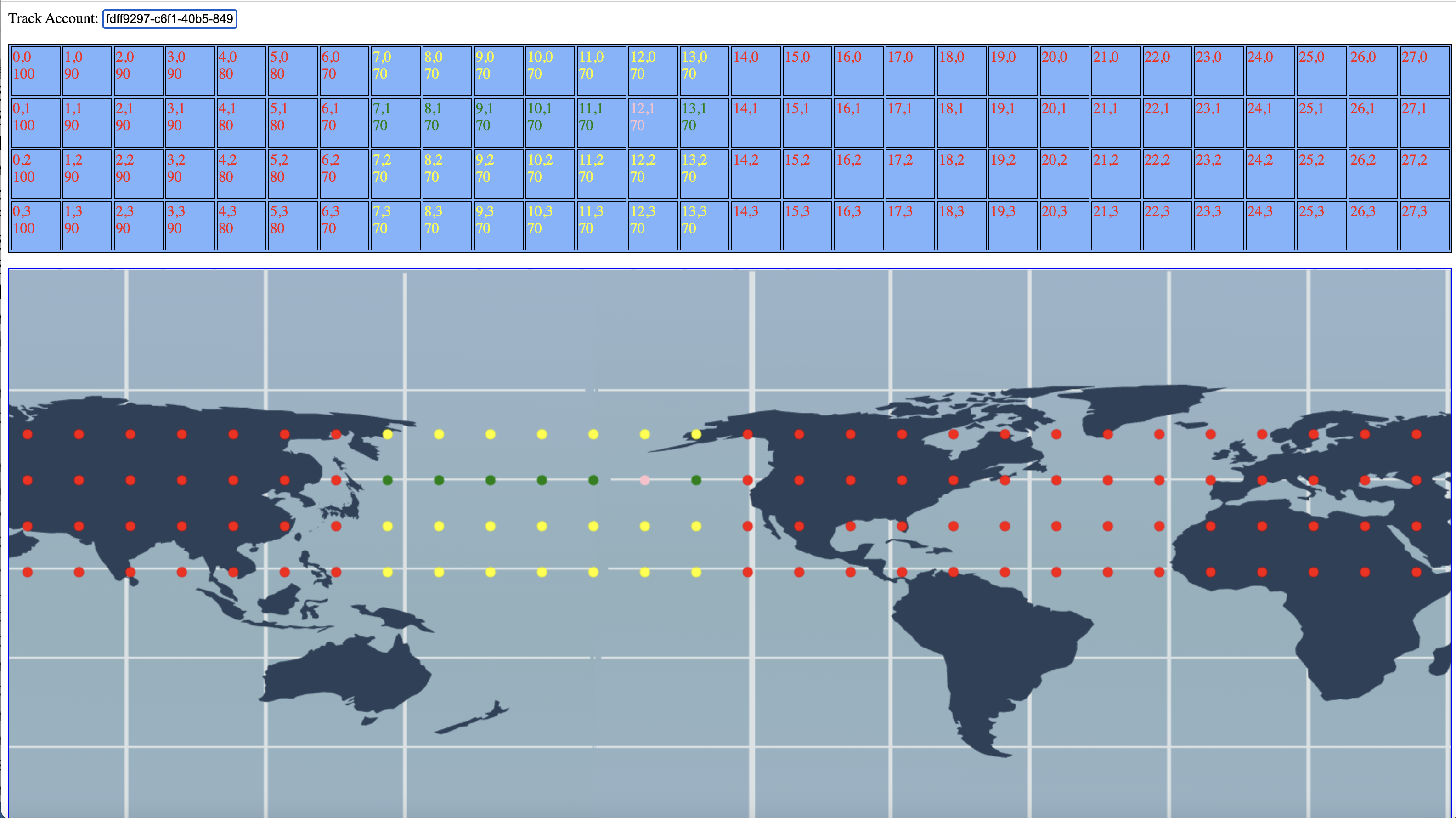}
        \caption{Simulation Web visualization screenshots of periods 1, 4, 6, and 8. Top: Torus cells, Bottom: Earth overlay. 
	Red node: outside of service area. Yellow node: inside service area. Green node: leader row node. Pink node: leader.}
\label{screenshot}
\end{figure}

A screenshot of the visualization can be seen in Figure~\ref{screenshot}.

%% file: evaluation.tex
\section{Communication Evaluation}\label{sec:evaluation}
We now take a closer look at the communication overhead of our solution.
In particular we are interested in the total number of messages generated
per transaction for different sizes of service areas.
We use a $28\times4$ grid of total nodes, and always use
all orbital planes but vary the range of satellites in a
given plane (x-range) that are in a service area. Varying
the x-range from 5 nodes to 10 nodes results in service
areas having 20 to 40 nodes. We use 70 threads in each
simulation process that represent an orbital plane. 

The evaluation involves executing 3 transactions, 2 account creations
and then a transfer between the newly created accounts
for each period, or rotational position of the earth. We define
28 rotational positions, meaning that after 28 periods the
satellites will be back in their original positions.

We keep the service area over the same geographic patch (Pacific Ocean)
so that new nodes enter and leave the area with every rotation.
For each configuration we do 10 full cycles of rotations around
the earth and then measure the number of messages sent in the system.
A message that is routed between nodes is counted as a new message
for each hop, as all communication is single hop according to the
LEO ISL setup previously described. The evaluation configuration
is summarized in Table~\ref{T:configuration}. 

\begin{table}[htbp]
        \caption{Evaluation Configuration.}
\begin{center}
\begin{tabular}{|l|l|}
\hline
Transactions & 841 \\
Migrations & 1120 \\
Earth Cycles & 10 \\
Cycle Periods & 28 \\
Service Area Nodes & $\{20,24,28,32,36,40\}$ \\
Transaction Commits & $\{16820,20184,23548,26912,30276,33640\}$ \\
\hline
\end{tabular}
\label{T:configuration}
\end{center}
\end{table}

\begin{figure}[htbp]
        \centerline{\includegraphics[scale=0.5]{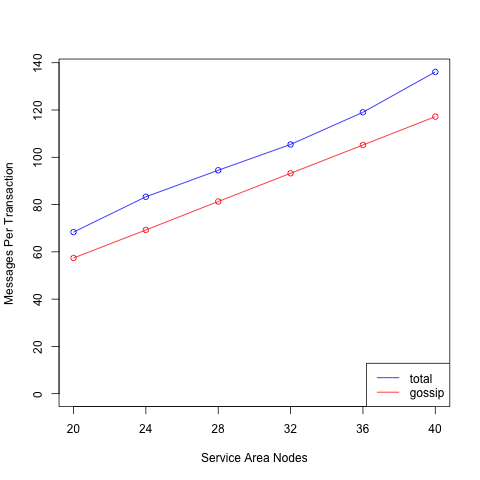}}
        \caption{Total messages and gossip messages (for transaction commits).}
\label{gossip}
\end{figure}

In Figure~\ref{gossip} we observe that the number of messages grows linearly
with the size of the service area and that the gossip messages dominate
the communication overhead for transaction processing. We note that the
transaction commits also grow linearly with service area nodes as per the
virtual synchrony design. Hence, adjusting the size of the service area
is an effective way to limit both communication and I/O processing overhead.

Table~\ref{T:messages} shows the proportion of different message types,
again highlighting the dominance of gossip messages. Only the
gossip messages change based on service area size. The {\it route\_execute}
message is sent if the sending node cannot execute the request, e.g.
it is not in a service area and is asked to execute a transaction.
The {\it sync\_blocks} message is a variant of this where a node is
asked by another node to sync state with its neighbor because it is
about to enter the service area. In a live deployment the node
would know internally when it needs to synchronize state as it
is aware of its orbital path. The {\it sync\_state} message
actually synchronizes the blockchain and its state as well
as in-memory state with its neighbor to the west to be able to participate in transaction
execution within the service area.

\begin{table}[htbp]
        \caption{Message proportions in $7\times 4$ grid service area.}
\begin{center}
\begin{tabular}{|l|l|}
\hline
\textbf{Message} & \textbf{Percent of total number} \\
 & \textbf{of messages} \\
\hline
gossip & 86\% \\
route\_execute & 9\% \\
execute\_transactions & 1\% \\
order\_transaction & 1\% \\
sync\_blocks & 1\% \\
sync\_state & 1\% \\
\hline
\end{tabular}
\label{T:messages}
\end{center}
\end{table}

\begin{figure}[htbp]
	\centering
        \includegraphics[scale=0.4]{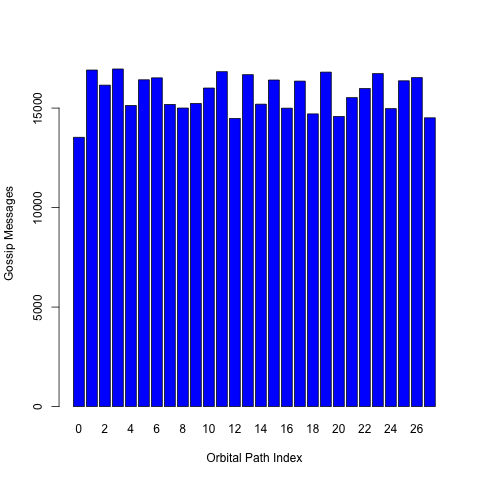}
        \includegraphics[scale=0.4]{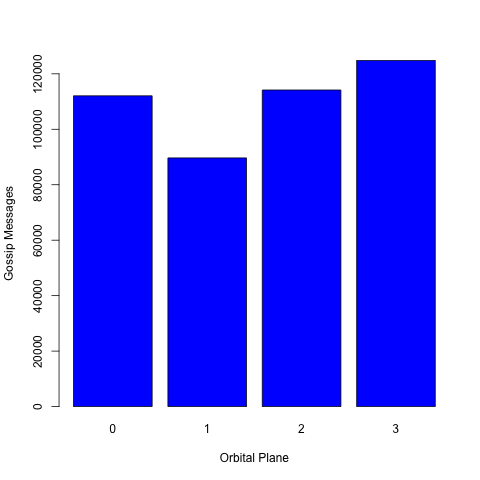}
	\caption{Gossip message distribution across orbital indices and orbital planes across all evaluation runs.}
\label{distribution}
\end{figure}

Figure~\ref{distribution} shows the gossip message distribution for satellites across orbital index and plane. The dip in plane 1 is because the leader originating gossip is in this plane (the leader row).

%% file: conclusion.tex
\section{Conclusions}\label{sec:conclusion}
We have demonstrated how a blockchain can be hosted
efficiently on-board LEO satellites to offer eventually
consistent guarantees for distributed transactions
and smart contracts. We believe that the importance and utility of such
data infrastructure will increase in the future
when satellite nodes are upgraded in compute and
storage capacity to avoid expensive round-trip costs
and to meet the stringent latency guarantees of 
3GPP NTN non-transparent direct-to-cell communication.  

There are many use cases for the data infrastructure proposed and evaluated in this work.  An example of an innovative application that uses the capability of such infrastructure is fast, on-demand authentication to a new MNO through an SNO via a smart
bandwidth contract that automates roaming.